\input amstex
\magnification=1200
\documentstyle{amsppt}
\NoRunningHeads
\pagewidth{142 mm}
\vcorrection{-15 mm}
\topmatter
\title	Darboux Transformation for Dirac Equations with $(1+1)$ potentials\endtitle
\author A V Yurov\endauthor
\address {\rm 236041, Theoretical Physics Department, Kaliningrad State
University, Al.Nevsky St., 14, Kaliningrad, Russia}
\endaddress
\email
yurov\@ freemail.ru
\endemail
\abstract
We study the Darboux transformation (DT) for Dirac equations with $(1+1)$ potentials.
 Exact
solutions for the adiabatic external field are constructed. The connection
between the exactly soluble Dirac (1+1) potentials and the soliton
solutions of the Davey--Stewartson equations is discussed.
\endabstract
\endtopmatter
\document

1. INTRODUCTION.

The Darboux transformation (DT) is a convenien way to construct a rich set of
integrable potentials of the steady--state Shr\"odinger equation in the single dimensional
case [1]. As  shown in [2], DT can be used to study the Dirac equation for a two--dimensional
fermion in an external scalar field $w(x)$. The aim of this work is the generalization
of DT for the Dirac equations with one--space--dimensional and non--stationary potentials
$u(t,x)$. In Sec. 2 we show that for a fermion
of transverse momenta $p\equiv p_y,$ $q\equiv p_z$ the four--dimensional
equation reduces to the Zakharov--Shabat equation. We demonstrate DT for this
equation and the connection with intertwining and supersymmetry algebra. The result
of the multiple DT (extended Crum law [3]) is set forth in Sec. 3. In Sec. 4 we
show the connection between the exactly solvable Dirac (1+1) potentials and the soliton
solutions of the Davey--Stewartson equations [4]. We also discuss the
reduction restriction problem.

2. DARBOUX TRANSFORMATION

Let us consider the four--dimensional  Dirac equation
$$(i\gamma^{\mu}\partial_{\mu}-\gamma^{\mu}A_{\mu}(t,x)+m)\Psi =0.\eqno(1)$$
We use the $\gamma$--matrix representation [2], [5]
$$
\gamma^0=\pmatrix
0&\sigma_2\\
\sigma_2&0
\endpmatrix,\qquad
\gamma^1=\pmatrix
0&-i\sigma_1\\
-i\sigma_1&0
\endpmatrix,
$$
$$
\gamma^2=\pmatrix
-iI&0\\
0&iI
\endpmatrix,\qquad
\gamma^3=\pmatrix
0&-i\sigma_3\\
-i\sigma_3&0
\endpmatrix,\eqno(2)
$$
and for $A^{\mu}(t,x)$ we have either:
$$A^{\mu}(t,x)=(0,0,A^2(t,x),A^3(t,x))^T,\eqno(3)$$
or
$$A^{\mu}(t,x)=(0,0,Q(t,x),cQ(t,x))^T,\qquad c=const.\eqno(4)$$
One can easily see that (3) and (4) are  solutions of Maxwell
equations.
Let $\Psi$ have the form:
$$
\Psi=\Phi(t,x)\exp(i(py+qz)),\qquad Im\,p=Im\,q=0.\eqno(5)
$$
Then $\Phi$ satisfies the Zakharov--Shabat equation:
$$
\Phi_t=J\Phi_x+U\Phi.\eqno(6)
$$
For the case (3):
$$
U=\pmatrix
0&\tilde A&0&m-\tilde B\\
\tilde A&0&\tilde B-m&0\\
0&\tilde B+m&0&\tilde A\\
-\tilde B-m&0&\tilde A&0
\endpmatrix,\qquad
J=\pmatrix
1&0&0&0\\
0&-1&0&0\\
0&0&1&0\\
0&0&0&-1
\endpmatrix,
\eqno(7)
$$
where $\tilde A=i(A^3-q)$, $\tilde B=i(p-A^2)$.

For the case (4) we get:
$$
\Phi =\pmatrix
A\Gamma\\
B\Gamma
\endpmatrix, \eqno(8)
$$
where
$$
A=\pmatrix
0&\alpha\\
\mu&0
\endpmatrix,\qquad
B=\pmatrix
0&\beta\\
{\mu}{\rho}&0
\endpmatrix,\qquad
\Gamma =\pmatrix
\psi (t,x)\\
\phi (t,x)
\endpmatrix,
$$
$$c=\frac {iq({\alpha}^2+{\beta}^2)+2m{\alpha}{\beta}}{ip({\alpha}^2+{\beta}^2)+m({\alpha}^2-{\beta}^2)},\qquad
\rho =\frac {i({\alpha}p+{\beta}q)+{\alpha}m}{i({\alpha}q-{\beta}p)+{\beta}m}.\eqno(9)$$
The condition $c^*=c$ gives us for $\alpha\equiv {\alpha_{_ R}}+i{\alpha_{_I}},$ $\beta\equiv {\beta_{_R}}+i{\beta_{_I}}$
$$\frac {\beta_{_I}}{\alpha_{_I}} =\frac {q{\alpha_{_R}}-p{\beta_{_R}}}{p{\alpha_{_R}}+q{\beta_{_R}}},\qquad
{(p{\alpha_{_R}}+q{\beta_{_R}})}^2({\alpha_{_R}}^2-{\alpha_{_I}}^2+{\beta_{_R}}^2)-{\alpha_{_I}}^2{(q{\alpha_{_R}}-p{\beta_{_R}})}^2+$$
$$+2m{\alpha_{_I}}({\beta_{_R}}^2+{\alpha_{_R}}^2)(p{\alpha_{_R}}+q{\beta_{_R}})=0.\eqno(10)$$
One of the nontrivial solutions of (10) is
$${\beta_{_I}}=0,\qquad{\beta_{_R}}=\frac {q{\alpha_{_R}}}{p},\qquad{\alpha_{_I}}^{\pm}={\frac {\alpha_{_R}}{p}}(m\pm \sqrt{m^2+p^2+q^2}).\eqno(11)$$

Substituting (8), (9) into (1) and taking (10) into account, we get a $2\times 2$
equation (6) for $\Gamma$ where $J=\sigma_3$,
$$
U=\pmatrix
0&u\\
v&0
\endpmatrix,
$$
$$u(t,x)=\lambda_1+{\lambda_2}Q(t,x),\qquad v(t,x)=\nu_1+{\nu_2}Q(t,x),$$
$$\lambda_1=\frac {m\beta +i(q{\alpha}-p{\beta})}{\mu},\qquad\lambda_2={\frac 1{\mu}}{\frac {({\alpha}^2+\beta^2)(q\alpha -p\beta -im\beta )}{ip(\alpha^2+\beta^2)+m(\alpha^2-\beta^2)}},$$
$$\nu_1=\frac {{\mu}(m^2+q^2+p^2)}{i(p\beta -q\alpha )-m\beta},\qquad\nu_2={\frac {i\mu}{\beta}}(1-c\rho ).\eqno(12)$$

Let $\chi_{_1}$ and $\chi_{_2}$ be $4\times 4$ (for the case (3)) or $2\times 2$ (for the case (4))
matrix solutions of equation (6). We define a matrix function $\tau_{_1}\equiv \chi_{1,x}\chi_{_1}^{-1}.$
It easy to see
that $\tau_{_1}$ satisfies the following  nonlinear equations:
$$\tau_{_{1,t}}=\sigma_3\tau_{_{1,x}}+[U,\tau_{_1}]+[\sigma_3,\tau_{_1} ]+U_x.\eqno(13)$$
Equation (6) is covariant with respect to DT:
$$\chi_{_2}[1]=\chi_{2,x}-\tau_{_1}\chi_{_2},\qquad U[1]=U+[\sigma_3,\tau_{_1}].\eqno(14)$$

It is necessary to choose the function $\chi_{_1}$ in such a way that the structure of the
matrix $U[1]$ be the same as the structure of the matrix $U$. This is the condition that we
call the reduction restriction (see Sec.4).

The transformation (14) allows us to constract a superalgebra, in just the
same way as the DT for the steady--state Shr\"odinger equation in a one--dimensional
case [6]. In order to do this we
introduce the following operators:
$$G^{(+)}={\frac {\partial}{\partial x}}-\tau_{_1},\qquad G^{(-)}={\frac {\partial}{\partial x}}+\tau_{_1}^+.\eqno(15)$$
Let us define new several operators as follow:
$$h\equiv G^{(-)}G^{(+)},\qquad h[1]\equiv G^{(+)}G^{(-)},\eqno(16)$$
$$ T\equiv {\frac {\partial}{\partial t}}-J{\frac {\partial}{\partial x}}-U,\qquad T[1]\equiv {\frac {\partial}{\partial t}}-J{\frac {\partial}{\partial x}}-U[1].\eqno(17)$$
It easy to see that
$$G^{(+)}T=T[1]G^{(+)},\qquad TG^{(-)}=G^{(-)}T[1],\qquad [h,T]=[h[1],T[1]]=0.\eqno(18)$$
The operators $h$ and $h[1]$ are the tipical  one--dimensional matrix Hamiltonians:
$$h={\frac {\partial^2}{\partial x^2}}+\rho_{_D}{\frac {\partial}{\partial x}}+V,\qquad h[1]={\frac {\partial^2}{\partial x^2}}+\rho_{_D}{\frac {\partial}{\partial x}}+V[1],\eqno(19)$$
$$\rho_{_D}=(\tau_{_1}^+-\tau_{_1})_{_D},\qquad V=-(\tau_{_{1,x}}+\tau_{_1}^+\tau_{_1}),\qquad V[1]=\tau_{_{1,x}}^+-\tau_{_1}\tau_{_1}^+,\eqno(20)$$
where $(\tau_{_1})_{_D}$ -- diagonal part of $\tau_{_1}$. It easy to see that the operators $q^{(\pm)}$, $H$
$$
q^{(+)}=\pmatrix
0&0\\
G^{(+)}&0
\endpmatrix,\qquad
q^{(-)}=\pmatrix
0&G^{(-)}\\
0&0
\endpmatrix,\qquad
H=\pmatrix
h&0\\
0&h[1]
\endpmatrix
\eqno(21)
$$
generate a supersymmetry algebra:
$$\{q^{(\pm)},q^{(\pm)}\}=[q^{(\pm)},H]=0,\qquad \{q^{(+)},q^{(-)}\}=H.\eqno(22)$$

Furthermore we restrict ourselves to studing  the case (4). The case (3) will
be considered in a separate work.

3. EXTENDED CRUM LAW

Let us consider $2N+1$ particular solutions of (6) with $\Gamma_k\equiv {(\psi_k,\phi_k )}^T,$ $k\le 2N,$ $\Phi\equiv ({\psi},{\phi})$:
$${\psi_{k,t}}+{\psi_{k,x}}+u(t,x)\phi_k =0,\qquad{\phi_{k,t}}-{\phi_{k,x}}+v(t,x)\psi_k =0.\eqno(23)$$

The following theorem is established:

THEOREM

Functions ${\psi}[N],$ ${\phi}[N]$ satisfy (23) with potentials $u[N]$ and $v[N]$ such that:
$${\psi}[N]=\frac {\Delta_1}{D},\qquad {\phi}[N]=\frac {\Delta_2}{D},\qquad u[N]=u+2{\frac {D_1}D},\qquad v[N]=v-2{\frac {D_2}D},\eqno(24)$$
where ${\Delta_{1,2}},$ $D_{1,2},$ $D$ are the following determinats $(\psi^{(N)}\equiv \frac {\partial^N\psi (t,x)}{\partial x^N}):$
$$
D=\vmatrix
\psi_{_1}^{(N-1)}&...&\psi_{_1}&\phi_{_1}^{(N-1)}&...&\phi_{_1}\\
\psi_{_2}^{(N-1)}&...&\psi_{_2}&\phi_{_2}^{(N-1)}&...&\phi_{_2}\\
.\\
.\\
.\\
\psi_{2N}^{(N-1)}&...&\psi_{2N}&\phi_{2N}^{(N-1)}&...&\phi_{2N}
\endvmatrix,
$$
$$
D_1=\vmatrix
\psi_{_1}^{(N)}&...&\psi_{_1}&\phi_{_1}^{(N-2)}&...&\phi_{_1}\\
\psi_{_2}^{(N)}&...&\psi_{_2}&\phi_{_2}^{(N-2)}&...&\phi_{_2}\\
.\\
.\\
.\\
\psi_{2N}^{(N)}&...&\psi_{2N}&\phi_{2N}^{(N-2)}&...&\phi_{2N}
\endvmatrix,
$$
$$
D_2=\vmatrix
\psi_{_1}^{(N-2)}&...&\psi_{_1}&\phi_{_1}^{(N)}&...&\phi_{_1}\\
\psi_{_2}^{(N-2)}&...&\psi_{_2}&\phi_{_2}^{(N)}&...&\phi_{_2}\\
.\\
.\\
.\\
\psi_{2N}^{(N-2)}&...&\psi_{2N}&\phi_{2N}^{(N)}&...&\phi_{2N}
\endvmatrix,
$$
$$
\Delta_1=\vmatrix
\psi^{(N)}&...&\psi&\phi^{(N-1)}&...&\phi\\
\psi_{_1}^{(N)}&...&\psi_{_1}&\phi_{_1}^{(N-1)}&...&\phi_{_1}\\
.\\
.\\
.\\
\psi_{2N}^{(N)}&...&\psi_{2N}&\phi_{2N}^{(N-1)}&...&\phi_{2N}
\endvmatrix,
$$
$$
\Delta_2=\vmatrix
\phi^{(N)}&\psi^{(N-1})&...&\psi&\phi^{(N-1)}&...&\phi\\
\phi_{_1}^{(N)}&\psi_{_1}^{(N-1})&...&\psi_{_1}&\phi_{_1}^{(N-1)}&...&\phi_{_1}\\
.\\
.\\
.\\
\phi_{2N}^{(N)}&\psi_{2N}^{(N-1})&...&\psi_{2N}&\phi_{2N}^{(N-1)}&...&\phi_{2N}
\endvmatrix.\eqno(24)
$$

To prove this
theorem we construct $N$ $2\times 2$ matrix functitons $\chi_k=(\Phi_{2k-1},\Phi_{2k}),\,1\le k\le N$:
$$
\chi_k=\pmatrix
\psi_{2k-1}&\psi_{2k}\\
\phi_{2k-1}&\phi_{2k}
\endpmatrix.
$$
These functions satisfy the matrix equation (6) with $J=\sigma_3$.

After N--time DT (14) we get $\chi[N]$ and $U[N]$ which satisfy (6). Let us write $\chi[N]$ as a series:
$$ \chi[N]=\chi^{(N)}-\sum_{i=1}^N A_i(t,x)\chi^{(N-i)},\qquad
A_i=\pmatrix
a_i&b_i\\
c_i&d_i
\endpmatrix.\eqno(25)
$$
Plugging (25) into  (23) we get (here $C_N^k=\frac {N!}{k!\,(N-k)!}$):
$$\sum_{k=0}^N C_N^kU^{(k)}\chi^{(N-k)}+ \sum_{k=1}^N[(A_{k,t}- \sigma_3A_{k,x})\chi^{(N-k)}- 2\sigma_3(A_k)_{_F}\chi^{(N-k+1)}]+$$ $$ +\sum_{k=1}^N \sum_{i=0}^{N-k} C_N^iA_kU^{(i)}\chi^{(N-k-i)}- U[N](\chi^{(N)}+ \sum_{k=1}^N A_k\chi^{(N-k)})=0,\eqno(26)$$
where $(A_k)_{_F}$ is the off--diagonal part of $A_k.$ Therefore
$$u[N]=u+2b_1,\qquad v[N]=v-2c_1.\eqno(27)$$
To compute $b_1$ and $c_1$ we take into account that $\chi_k[N]=0$ if $k\le N$,
therefore we get a system of $2N$ equations as follows:
$$\psi_i^{(N)}=\sum_{n=1}^N (a_n\psi_i^{(N-n)}+b_n\phi_i^{(N-n)}),\qquad
\phi_i^{(N)}=\sum_{n=1}^N (c_n\psi_i^{(N-n)}+d_n\phi_i^{(N-n)}),\eqno(28)$$
$i=1,...,N.$ Using Kramer's formulae, we get  (24).

If (see(12)) $\mu= i(\beta^*-c\alpha^*)^{-1}$ then $Im\,\nu_2=Im\,\lambda_2=0$. The constants
$\lambda_1$ and $\nu_1$ may be annihilated by the standard gauge $U(1)$ transformations.
Using freedom in our choice of parameters, let us assume that $v(t,x)=\kappa u(t,x) , \kappa=\pm 1.$
Now we can  supplement (24) with the  transformations law for $\chi_{_1}$ ($N=1$):
$$\psi_{_1}[1]=\frac {\phi_{_2}}{\Delta},\qquad \psi_{_2}[1]=\frac {\kappa\phi_{_1}}{\Delta},\qquad \phi_{_1}[1]=\frac {\kappa\psi_{_2}}{\Delta},$$
$$\phi_{_2}[1]=\frac {\psi_{_1}}{\Delta},\qquad \Delta=\vmatrix
\psi_{_1}&\psi_{_2}\\
\phi_{_1}&\phi_{_2}
\endvmatrix.\eqno(29)
$$
It is necessary to require that after N--time DT the following reduction restriction will be true:
$$v[N]=\kappa u[N].\eqno(30)$$
In the general case we do not have an algorithm  allowing us to keep (30) in all the steps  of
DT. However, it is possible by the introduction of the	so called $binary$ $DT$ which allows one
to preserve the reduction restriction (30) [7].

Let us consider a closed 1--form
$$d\Omega= dx\,\zeta\chi +dt\,\zeta\sigma_3\chi,\qquad \Omega\equiv \int d\Omega\eqno(31)$$
where a $2\times 2$ matrix function $\zeta$ solves the equation:
$$\zeta_t =\zeta_x\sigma_3 -\zeta U.\eqno(32)$$
We shall apply the DT for (6). One can verify by  immediate substitution
that (32) is covariant with respect to the transform if
$$\zeta[+1]=\Omega (\zeta,\chi)\chi^{-1}.\eqno(33)$$
Now we can   alternatively affect $U$, by the following transformaton:
$$U[+1,-1] =U+[\sigma_3,\chi\Omega^{-1}\zeta].\eqno(34)$$
It may be shown that
$$\chi[+N,-N] =\chi -\sum_{k=1}^N \theta_k\Omega(\zeta_k,\chi),\qquad \zeta[+N,-N] =\zeta -\sum_{k=1}^N \Omega(\zeta,\chi_k)s_k,\eqno(35)$$
where $\theta_k$ and $s_k$ may be found from  the following equations:
$$\sum_{k=1}^N \theta_k\Omega(\zeta_k,\chi_i) =\chi_i,\qquad \sum_{k=1}^N \Omega(\zeta_i,\chi_k)s_k =\zeta_i.\eqno(36)$$
The transformation:
$$U[+N,-N] =U +\sum_{i,k=1}^N [\sigma_3,\theta_i\Omega(\zeta_k,\chi_i)s_k]\eqno(37)$$
is the forementioned $binary$ DT.

Let $v=\kappa u^*$ (see Sec. 4; in this section $u$ and $v$ are real so the condition
is equivalent to (30)), then $U[+N,-N]$ will  satisfy the  reduction restriction
if we choose $\zeta_k$ and $\chi_k$ such that:
$$\zeta_k =\chi_kR,\qquad R=diag(1,-\kappa).\eqno(38)$$
The solution that  follows from (34) has the form
$$u[+1,-1]=u+\frac {2\kappa(\psi_{_2}\phi_{_1}^*\theta_{12}^* +\psi_{_1}\phi_{_2}^*\theta_{12}-\psi_{_1}\phi_{_1}^*\theta_{22}-\psi_{_2}\phi_{_2}^*\theta_{11})} {\theta_{11}\theta_{22}-\mid \theta_{12}\mid^2},\eqno(39)$$
$$\theta_{ik}=\int dx\,(\psi_i^*\psi_k-\kappa \phi_i^*\phi_k)+dt\,(\psi_i^*\psi_k+\kappa \phi_i^*\phi_k).$$
Note that the square of the absolute value $u[+1,-1]$ is expressed by the compact formula:
$$\mid u[+1,-1]\mid^2=\mid u\mid^2-\kappa ({\frac {\partial^2}{\partial t^2}} -{\frac {\partial^2}{\partial x^2}})\ln(\theta_{11}\theta_{22}-\mid \theta_{12}\mid^2).\eqno(40)$$

DT allow us to constract a rich set of the exact solutions of the Dirac equations with
(1+1) potentials. In conclusion of this section we consider decreasing in $t,x \to \pm\infty$
fields (adiabatic engaging and turning-off). Let $u=v=0$. The particular
solutions of (23) are:
$$\psi_k=A_k\,e^{\omega(t-x)}+B_k\,e^{\omega(x-t)},\qquad
\phi_k=C_k\,e^{\lambda(t+x)}+D_k\,e^{-\lambda(t+x)},\eqno(41)$$
where $k=1,2$; $A$, $B$, $C$, $D$, $\omega$ and $\lambda$ are real constants. After
DT (14) under condition (12) we get:
$$Q(t,x)=\frac {1}{\mu_1\,\cosh(at+bx+\delta_1)+\mu_2\,\cosh(bt+ax+\delta_2)},\eqno(42)$$
$$u(t,x)=\xi\,Q(t,x),\qquad v(t,x)=-{\frac {1+c^2}{\xi^2}}u(t,x),\eqno(43)$$
$$\xi=\mid \alpha c -\beta\mid^2=4\omega(A_1B_2-A_2B_1),\,\, a=\omega+\lambda,\,\, b=\lambda-\omega,\,\, \mu(\beta^*-c\alpha^*)=i,$$
$$1+c^2=4\lambda\xi(C_2D_1-C_1D_2),\qquad \omega(A_1B_2-A_2B_1)=\lambda(D_1C_2-D_2C_1),$$
$$\mu_1=\sqrt {(A_1C_2-A_2C_1)(B_1D_2-B_2D_1)},\,\, \mu_2=\sqrt {(B_1C_2-B_2C_1)(A_1D_2-A_2D_1)},$$
$$\delta_1={\frac {1}{2}}\ln{\frac {A_1C_2-A_2C_1}{B_1D_2-B_2D_1}},\qquad \delta_2={\frac {1}{2}}\ln{\frac {B_1C_2-B_2C_1}{A_1D_2-A_2D_1}},$$
where $\alpha,\beta$ satisfy (11). The potential (42) is a localized impulse, which
decreases in $t \to \pm\infty$. A particular solution of the Dirac equation may be
calculated using (5), (8) from $(\psi_{1,2}[1]$,$\phi_{1,2}[1])^T$, where $\psi_{1,2}[1]$
and $\phi_{1,2}[1]$ are defined by (29). This bispinor describes the quazi--stationary
state of a fermion. The fermion is free along the $y$ and $z$ and restrained  along the $x$--axis.

4. DAVEY-STEWARTSON EQUATIONS.

It is shown in [2]  that a class of the exactly soluble  Dirac one--dimensional potentials corresponds to a
soliton solutions of the MKdV equation, just as certain Schr\"odinger potentials
are solutions of the KdV equation. In this section we show that there exist the analogous connection
between exactly soluble  Dirac (1+1) potentials and soliton solutions of  the
Davey--Stewartson equations (DS).

The DS equations  appear as the commutation condition of the two operators [7], [8]:
$$[T_1,T_2]=0,\eqno(44)$$
where $T_1\equiv T$ (see (17)) and $T_2$ is given by the formula:
$$T_2=i{\frac {\partial}{\partial y}}+2\sigma_3{\frac {\partial^2}{\partial x^2}}+2U{\frac {\partial}{\partial x}}+U_x+\sigma_3U_t+A,\eqno(45)$$
$$A=diag(A_1,A_2),\qquad v=\kappa u^*,$$
$$A_1=-\kappa\mid u\mid^2+{\frac {1}{2i}}l_{(+)}F,\qquad A_2=\kappa\mid u\mid^2+{\frac {1}{2i}}l_{(-)}F.$$
Here $l_{(\pm)}={\frac {\partial}{\partial t}}\pm {\frac {\partial}{\partial x}}$ and
$F$ is a pure imaginary function. The operator $T_2$ is also covariant with respect
to DT, what allows one to get infinite sets of exact solutions of DS, for example soliton solutions, exponentially
localized on the plane -- the dromions (42) [7], [8]. The DS equations contain
two fields: $u(y,t,x)$ -- the amplitude wavetrain function and $S(y,t,x)\equiv -iF_x$ --
the amplitude function of slowly changing in $t$ (space variable!) and $x$  mean
field:
$$iu_y+u_{xx}+u_{tt}-2\kappa\mid u\mid^2u+Su=0,\qquad l_{(+)}l_{(-)}S=-4\kappa (\mid u\mid^2)_{xx}.\eqno(46)$$
The dromions $S$--component is constant along the two ortogonal directions and moves
with the constant velocity [8].

In Sec.3 we have shown that the one--dromion solution corresponds to adiabatic engaging
and turning-off electro-magnetic field. Boity, Leon, Martina and Pempinelli [8]
showed that two dromions scatter against  each other with the soliton phase shift, therefore the N--dromions
solution corresponds to a adiabatic external field, too. A many--dromion solution may be
obtained by the N--time DT. For this purpose we transform the $LA$--pair ($T_1$ and $T_2$).
Introducing new variables $x=p+q$ and $t=p-q$ we exclude a field $F$ from (45).
The matrix $A$ takes the form
$$
A=\pmatrix
\kappa\int_{-\infty}^q dq\,\frac {\partial \mid u\mid^2}{\partial p} +g_1(p,y)&0\\
0&\kappa\int_{-\infty}^p dp\,\frac {\partial \mid u\mid^2}{\partial q} +g_2(q,y)
\endpmatrix,\eqno(47)
$$
where $g_1(p,y)$ and $g_2(q,y)$ are arbitrary functions. We suppose that $u(p,q,y)$
belongs to a Schwarz space $L$. Then a nonlocal flow is given as follows:
$$S=2\kappa\mid u\mid^2+({\int_{-\infty}^q dq\,\frac {\partial}{\partial p}}+{\int_{-\infty}^p dp\,\frac {\partial}{\partial q}})\mid u\mid^2 +g_1-g_2.\eqno(48)$$
In the new variables
$$
T_1=\pmatrix
-\frac {\partial}{\partial q}&u\\
\kappa u^*&\frac {\partial}{\partial p}
\endpmatrix,\eqno(49)
$$
therefore for $u \in L$, the equation $T_1\chi =0$ gives at infinity for components of $\chi$
 the following condition:
$$\psi_k=\psi_k(p,y),\qquad \phi_k=\phi_k(q,y),\qquad k=1,2.\eqno(50)$$
The second equation $T_2\chi =0$ gives four nonstationary one--dimensional
Schr\"odinger equations
$$i\psi_{k,y}+{\frac {1}{2}}\psi_{k,pp}+v_1\psi_k=0\eqno(51)$$
$$i\phi_{k,y}-{\frac {1}{2}}\phi_{k,qq}+v_2\phi_k=0\eqno(52)$$
where
$$v_1(p,y)=g_1+\kappa\int_{-\infty}^{+\infty} dq\,\frac {\partial \mid u\mid^2}{\partial p},\qquad
v_2(q,y)=g_2+\kappa\int_{-\infty}^{+\infty} dp\,\frac {\partial \mid u\mid^2}{\partial q}.\eqno(53)$$
For the case of a dromion (42) we have
$$v_1=4\omega^2 sec^2f_1,\qquad v_2=-4\lambda^2 sec^2f_2,$$
$$f_1=2\omega(p-2ay)+{\frac {1}{2}}\ln{\frac {A_0}{b_0}},\qquad
f_2=2\lambda(q+2by)+{\frac {1}{2}}\ln{\frac {C_0}{D_0}},\eqno(54)$$
where $a$, $b$, $(A, B, C, D)_0$ are real constants. So the dromion sits on two plane
solitons that correspond to the one--level reflectionless potentials. They  may
be easily obtained by the standard DT for the Schr\"odinger equation [1] on zero background:
$$\psi_k[1]=\frac {{\frac {\partial\psi_k}{\partial p}}\psi_0-{\frac {\partial\psi_0}{\partial p}}\psi_k}{\psi_0},\,\,\,
\phi_k[1]=\frac {{\frac {\partial\phi_k}{\partial q}}\phi_0-{\frac {\partial\phi_0}{\partial q}}\phi_k}{\phi_0},$$
$$v_1[1]=-2{\frac {\partial^2}{\partial p^2}}\ln\psi_0,\qquad v_2[1]=2{\frac {\partial^2}{\partial q^2}}\ln\phi_0.\eqno(55)$$
The background eigenfunctions are choosen such that:
$$\psi_0(p,y)=2{\sqrt {A_0B_0}}\,\cosh f_1\,e^{i\theta},\eqno(56)$$
$$\phi_0(q,y)=2{\sqrt {C_0D_0}}\,\cosh f_2\,e^{i\vartheta},\eqno(57)$$
where
$$\theta=2[(\omega^2-a^2)y+ap],\qquad \vartheta=2[(b^2-\lambda^2)y+bq],\eqno(58)$$
and the new coefficients $(A, B, C, D)_0$ satisfy the conditions:
$$A_0(B_1C_2-B_2C_1)=B_0(A_1C_2-A_2C_1),\qquad C_0(A_1D_2-A_2D_1)=D_0(B_1C_2-B_2C_1).$$

For the second DT it is necessary to switch to the two--level potentials in equations (51), (52).
Let $\varpi >\omega$, $\varrho >\lambda$ then two "linearly independent" solutions that
correspond to the eigenvalues $2\varpi$ and $2\varrho$ have the form:
$$\psi_{-1}^{(+)}=A_{-1}\,e^{2\varpi(p-2\alpha y)+i\theta_{-1}},\qquad
\psi_{-1}^{(-)}=B_{-1}\,e^{-2\varpi(p-2\alpha y)+i\theta_{-1}}\eqno(59)$$
for (51) and
$$\phi_{-1}^{(+)}=C_{-1}\,e^{2\varrho(q+2\beta y)+i\vartheta_{-1}},\qquad
\phi_{-1}^{(-)}=D_{-1}\,e^{-2\varrho(q+2\beta y)+i\vartheta_{-1}}\eqno(60)$$
for (52), where $\theta_{-1}$ and $\vartheta_{-1}$ differ from $\theta$ and
$\vartheta$ by	substitution $(a,b,\lambda,\omega)\to (\alpha,\beta,\varrho,\varpi).$
We transform $\psi_{-1}^{(\pm)}$ and $\phi_{-1}^{(\pm)}$ by the formula (55) with
$\psi_0$ and $\phi_0$ accordingly. Then we will build the  new	support function:
$$\psi_{-1}[1]={\psi_{-1}^{(+)}}[1]-{\psi_{-1}^{(-)}}[1],\qquad
\phi_{-1}[1]={\phi_{-1}^{(+)}}[1]-{\phi_{-1}^{(-)}}[1].\eqno(61)$$
For simplicity let us choose $A=B=C=D=1.$ As a result we have:
$$\psi_{-1}[1]=({\frac {(\varpi-\omega)\cosh(\xi_0+\xi_1)+(\varpi+\omega)\cosh(\xi_0-\xi_1)}{\cosh\,\xi_0}}+i(\alpha-a)\sinh\,\xi_0)e^{i\theta_{-1}},\eqno(62)$$
$$\phi_{-1}[1]=({\frac {(\varrho-\lambda)\cosh(\eta_0+\eta_1)+(\varrho+\lambda)\cosh(\eta_0-\eta_1)}{\cosh\,\eta_0}}+i(\beta-b)\sinh\,\eta_0)e^{i\vartheta_{-1}},\eqno(63)$$
where $\xi_0=2\omega(p-2ay), \xi_1=2\varpi(p-2ay), \eta_0=2\lambda(q+2by), \eta_1=2\varrho(q+2by).$
If one puts $\alpha=a$, $\beta=b$ then (62), (63) exactly coincide with the supporting
functions that generate the two--level potential with respect to the Darboux transformations [8].

Now it is necessary to define four functions $\psi_k[1]$, $\phi_k[1]$ $(k=1,2)$
which are the solutions of (51), (52) with the potentials (54). The basic problem at this
step is the independence of the wronskians
$$\vmatrix
{\frac {\partial \psi_{_2}}{\partial p}}&{\frac {\partial \psi_{_1}}{\partial p}}\\
\psi_{_2}&\psi_{_1}
\endvmatrix,\qquad
\vmatrix
{\frac {\partial \phi_{_2}}{\partial q}}&{\frac {\partial \phi_{_1}}{\partial q}}\\
\phi_{_2}&\phi_{_1}
\endvmatrix,
$$
with respect to the space coordinates, what is necessary for the solvability of the
reduction restriction equation $v[2]=\kappa\,u^*[2].$ For this purpose it is enough
to choose:
$$\psi_k[1]=A_{k,-1}(\varpi-\omega\,\tanh\,\xi_0)\psi_{-1}^{(+)}+B_{k,-1}(\varpi+\omega\,\tanh\,\xi_0)\psi_{-1}^{(-)},\eqno(64)$$
$$\phi_k[1]=C_{k,-1}(\varrho-\lambda\,\tanh\,\eta_0)\phi_{-1}^{(+)}+D_{k,-1}(\varrho+\lambda\,\tanh\,\eta_0)\phi_{-1}^{(-)},\eqno(65)$$
with the constants that satisfy the relation:
$$\varpi(\omega^2-\varpi^2)(B_{1,-1}A_{2,-1}-B_{2,-1}A_{1,-1})=\kappa\,\varrho(\lambda^2-\varrho^2)(C_{1,-1}D_{2,-1}-C_{2,-1}D_{1,-1}).\eqno(66)$$
Therefore we choose $\psi_{-2}[2],$ $\phi_{-2}[2]$ as the wave functions that
generate a three--level potential (via DT) and determine the corresponding functions
$\psi_k[2]$ and $\phi_k[2]$. One may repeat this procedure and realize the third DT.

ACNOWLEDGMENTS

The author would like to express his cordial thanks to M. Rudnev for his attention to this work and for valuable discussion.
This work was partially sponsored by the RFFI under Grant No. 91-01-01789.

\enddocument